\documentclass{webofc}
\usepackage[varg]{txfonts}   
\usepackage{tikz-feynman}
\usepackage{float}
\usepackage{subcaption}
\usepackage{graphicx}
\usepackage{amsfonts}
\usepackage{amsmath}
\usepackage{amssymb}
\begin{document}
\title{Thermalization and quark production in spatially \\homogeneous systems of gluons}
%
%

\author{\firstname{Sergio} \lastname{Barrera Cabodevila}\inst{1}\fnsep\thanks{\email{sergio.barrera.cabodevila@usc.es}} \and
        \firstname{Carlos A.} \lastname{Salgado}\inst{1, 2}\fnsep\thanks{\email{carlos.salgado@usc.es}} \and
        \firstname{Bin} \lastname{Wu}\inst{1}\fnsep\thanks{\email{bin.wu@usc.es}}
}

\institute{Instituto Galego de F\'isica de Altas Enerx\'ias IGFAE, Universidade de Santiago de Compostela,
E-15782 Galicia-Spain
\and
           Axencia Galega de Innovaci\'on (GAIN), Xunta de Galicia, Galicia-Spain
          }

\abstract{%
  We first assemble a full set of the Boltzmann Equation in Diffusion Approximation (BEDA) for studying thermalization/hydrodynamization and quark production in out of equilibrium systems. We then discuss thermalization and the production of three flavors of massless quarks in spatially homogeneous systems initially filled only with gluons. A complete parametric understanding for thermalization and quark production is obtained for both initially very dense or dilute systems, which are complemented by detailed numerical simulations. For initial distributions more relevant for heavy-ion collisions, the complete thermal equilibration is found to be significantly delayed by considering quark production.
}
\maketitle
\section{Introduction}
\label{intro}

Different tools have been proposed in order to study and comprehend the behavior of QCD bulk matter in heavy-ion collisions during the different stages of its short life. It is well known that right after the collision an out of equilibrium and highly populated system of gluons will be produced, as well as the system has time to exhibit an hydrodynamic behavior. In order to understand how thermalization/hydronamization is produced, we present the Boltzmann Equation in Diffusion Approximation (BEDA), and apply it to the simple case of spatially homogeneous systems of gluons as a testing of its consistency, as detailed in ref. \citep{Cabodevila:2023htm}.

\section{The Boltzmann Equation in Diffusion Approximation}
\label{sec-1}
The QCD Boltzmann equation at leading order reads
\begin{equation}
        (\partial_t {+ \mathbf{v} \cdot {\nabla_{\mathbf{x}}}}) f^a = C^a_{2 \leftrightarrow 2}[f] + C^a_{1 \leftrightarrow 2} [f].
        \label{Eq:Boltzmann}
\end{equation}
Where $f^a$ is the distribution function for particle $a=\lbrace g,q,\bar{q} \rbrace$, and $C^a_{2 \leftrightarrow 2}$ and $C^a_{1 \leftrightarrow 2}$ are the collision kernels for the $2 \leftrightarrow 2$ and $1 \leftrightarrow 2$ processes, respectively. In this study we will focus on spatially homogeneous systems. Thus, the spatial derivative in Eq. \eqref{Eq:Boltzmann} will vanish. Also, we will neglect any angular dependence, such that $f^a(\mathbf{p}) = f^a(p)$. 

\subsection{The $2 \leftrightarrow 2$ kernel}
\label{2to2}

The $2 \leftrightarrow 2$ kernel can be expressed in diffusion approximation as a Fokker-Planck term plus an additional source term \cite{Mueller:1999fp, Blaizot:2014jna},
\begin{equation}
\begin{gathered}
C^a_{2\leftrightarrow2}=\frac{1}{4}\hat{q}_a(t)\nabla_{\mathbf{p}} \cdot\left[  \nabla_{\mathbf{p}}f^a  + \frac{{\mathbf{v}}}{T^*(t)}f^a(1+\epsilon_a f^a)\right]+\mathcal{S}_a \equiv 
			\left(
            \begin{array}{c}
            \scalebox{0.28}{
               \begin{tikzpicture}
                    \begin{feynman}
                    
                    \vertex (i1);
                    \vertex[right=1.5 of i1] (i2);
                    \vertex[right=0.75 of i2] (i3);
                    \vertex[right=0.75 of i3] (i4);
                    \vertex[below=1.5 of i1] (j1);
                    \vertex[right=1.5 of j1] (j2);
                    \vertex[right=1.5 of j2] (j3);
                    \vertex[above=0.75 of i4] (e1);
                    
                    \diagram* {
                        (i1) -- [gluon] (i2) -- [gluon] (i4),
                        (j1) -- [gluon] (j2) -- [gluon] (j3),
                        (i2) -- [gluon] (j2),
                        
                    };
                    
                    \end{feynman}
                \end{tikzpicture}} \\

                \scalebox{0.28}{
                \begin{tikzpicture}
                    \begin{feynman}
                    
                    \vertex (i1);
                    \vertex[right=1.5 of i1] (i2);
                    \vertex[right=0.75 of i2] (i3);
                    \vertex[right=0.75 of i3] (i4);
                    \vertex[below=1.5 of i1] (j1);
                    \vertex[right=1.5 of j1] (j2);
                    \vertex[right=1.5 of j2] (j3);
                    \vertex[above=0.75 of i4] (e1);
                    
                    \diagram* {
                        (i1) -- [fermion] (i2) -- [fermion] (i4),
                        (j1) -- [fermion] (j2) -- [fermion] (j3),
                        (i2) -- [gluon] (j2),
                        
                    };
                    
                    \end{feynman}
                \end{tikzpicture}} \;

                \scalebox{0.28}{
                \begin{tikzpicture}
                    \begin{feynman}
                    
                    \vertex (i1);
                    \vertex[right=1.5 of i1] (i2);
                    \vertex[right=0.75 of i2] (i3);
                    \vertex[right=0.75 of i3] (i4);
                    \vertex[below=1.5 of i1] (j1);
                    \vertex[right=1.5 of j1] (j2);
                    \vertex[right=1.5 of j2] (j3);
                    \vertex[above=0.75 of i4] (e1);
                    
                    \diagram* {
                        (i1) -- [fermion] (i2) -- [fermion] (i4),
                        (j1) -- [gluon] (j2) -- [gluon] (j3),
                        (i2) -- [gluon] (j2),
                        
                    };
                    
                    \end{feynman}
                \end{tikzpicture}}
\end{array}             \right)                       
                +
                \left(
				\begin{array}{c}
                \scalebox{0.28}{
                \begin{tikzpicture}
                    \begin{feynman}
                    
                    \vertex (i1);
                    \vertex[right=1.5 of i1] (i2);
                    \vertex[right=0.75 of i2] (i3);
                    \vertex[right=0.75 of i3] (i4);
                    \vertex[below=1.5 of i1] (j1);
                    \vertex[right=1.5 of j1] (j2);
                    \vertex[right=1.5 of j2] (j3);
                    \vertex[above=0.75 of i4] (e1);
                    
                    \diagram* {
                        (i1) -- [fermion] (i2) -- [gluon] (i4),
                        (j1) -- [gluon] (j2) -- [fermion] (j3),
                        (i2) -- [fermion] (j2),
                        
                    };
                    
                    \end{feynman}
                \end{tikzpicture}} \\

                \scalebox{0.28}{
                \begin{tikzpicture}
                    \begin{feynman}
                    
                    \vertex (i1);
                    \vertex[right=1.5 of i1] (i2);
                    \vertex[right=0.75 of i2] (i3);
                    \vertex[right=0.75 of i3] (i4);
                    \vertex[below=1.5 of i1] (j1);
                    \vertex[right=1.5 of j1] (j2);
                    \vertex[right=1.5 of j2] (j3);
                    \vertex[above=0.75 of i4] (e1);
                    
                    \diagram* {
                        (i1) -- [gluon] (i2) -- [fermion] (i4),
                        (j3) -- [fermion] (j2) -- [gluon] (j1),
                        (j2) -- [fermion] (i2),
                        
                    };
                    
                    \end{feynman}
                \end{tikzpicture}}
				
				\end{array} \right)
\\
\mathcal{S}_q =\frac{2\pi\alpha_s^2  C_F^2 \mathcal{L}}{ p}\bigg[\mathcal{I}_c f (1-F ) - \bar{\mathcal{I}}_c F ( 1 + f ) \bigg], \quad
\mathcal{S}_{\bar{q}}=\mathcal{S}_q|_{F\leftrightarrow\bar{F}},\quad\mathcal{S}_g=-\frac{N_f}{2C_F}({\mathcal{S}_q+\mathcal{S}_{\bar{q}}}).
\end{gathered}
\end{equation}

We have renamed $f^g \equiv f$, $f^q \equiv F$ and $f^{\bar{q}} \equiv \bar{F}$ and defined $\mathcal{L}\equiv\ln\frac{\langle p_t^2\rangle}{m_D^2}$. Here, $\hat{q}$ is the quenching parameter, $m_D^2$ is the screening mass, $T_* \propto \hat{q} / (2 m_D^2)$ is the effective temperature and $\mu_*\equiv T_* \ln \frac{\mathcal{I}_c}{\bar{\mathcal{I}}_c}$ is the quark chemical potential\footnote{$\mu_*$ is defined for each quark flavour. If there is quark-antiquark symmetry, then $\mu_*=0$.}.

When this kernel is the only interaction included in the Eq. \eqref{Eq:Boltzmann}, a divergence appear if the system of initial gluons is highly occupied. This is interpreted as the onset of a Bose-Einstein condensate \cite{Blaizot:2013lga}, which can be studied numerically by adding appropriate boundary conditions to the simulation \cite{Blaizot:2014jna}. This boundary allows to save the extra gluons into the Bose-Einstein condensate as $\dot{n}_c \propto \lim_{p\rightarrow 0} pf - T_*$.

\subsection{The $1 \leftrightarrow 2$ kernel}

On the other side, the $1 \leftrightarrow 2$ is computed in the deep LPM regime \cite{Baier:1996kr, Arnold:2008zu}. The general expression for this kernel is more complicated than the previous one
\begin{align}
\begin{split}
        C^a_{1\leftrightarrow2} = \int_0^1\frac{dx}{x^3}\sum\limits_{b,c}\bigg[&\frac{\nu_c}{\nu_a}C^c_{ab}(\mathbf{p}/{x};\mathbf{p},\mathbf{p}(1-x)/x) -\frac{1}{2}C^a_{bc}(\mathbf{p};x \mathbf{p},(1-x)\mathbf{p})\bigg].
\end{split}
\end{align}
$C^a_{bc}$ represents the product of the splitting rate for the process $a\leftrightarrow bc$ and its correspondent statistical factor given by the distribution function of the particles involved. Notice that the sum over $b,c$ is only valid if the splitting/merging process $a\leftrightarrow bc$ is allowed by the QCD interaction vertex for three particles. That is, this kernel will only include $g \leftrightarrow gg$, $g \leftrightarrow q \bar{q}$ and $q \leftrightarrow g q$ (and the ones switching $q \leftrightarrow \bar{q}$).

\subsection{Rapid thermalization of the soft sector}
\label{Sec:RapidThermalization}

In order to see what happens with the condensate after $1 \leftrightarrow 2$ kernel is introduced, we should notice that the dominant contribution to the evolution at small values of $p$ is given by the $1 \leftrightarrow 2$ kernel. In particular, for a system initially occupied by gluons, the $g \rightarrow gg$ will be the dominant contribution to the gluon production, meanwhile the $g \rightarrow q\bar{q}$ will dominate the quark/antiquark production. This fact also tells us that the distributions of quarks/antiquarks and gluons will fill a thermal distribution really quickly up to some soft momentum scale $p_s$.
\begin{align}
        f^g(p) &\approx \frac{T_*}{p} &\mathrm{for} \, \, p \lesssim p_{s,g} \equiv (\hat{q}_A m_D^4 t^2/2)^\frac{1}{5}, \\
        f^q(p) &\approx \frac{1}{e^{-\frac{\mu_*}{T_*}} + 1} &\mathrm{for}  \, \, p \lesssim p_{s,q} \equiv [\alpha_s C_F\pi(\mathcal{I}_c+\bar{\mathcal{I}}_c)t]^\frac{2}{5}\hat{q}_F^\frac{1}{5}.
\end{align}
This implies that the variation of the particles of the Bose-Einstein condensate $\dot{n}_c \propto \lim_{p\rightarrow 0} pf - T_*$, as defined in Section \ref{2to2}, will be exactly zero for gluons.

\section{The underpopulated scenario}

Now, let us review the behavior of a system initially populated with gluons, similar to the scenario that Color Glass Condensate predicts right after the collision. In this case, and as was already identified for the pure gluon system \citep{BARRERACABODEVILA2022137491}, thermalization is achieved in three different stages: overheating, cooling/overcooling and reheating.

\begin{figure}[t]
\centering
\begin{subfigure}[b]{0.32\textwidth}
	\centering
	\includegraphics[width=\textwidth]{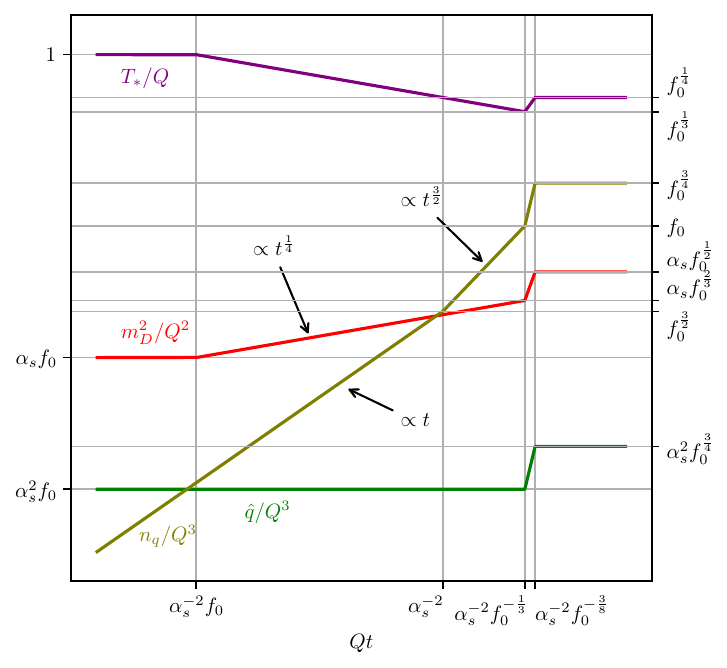}
\end{subfigure}
\begin{subfigure}[b]{0.32\textwidth}
	\centering
	\includegraphics[width=1.03\textwidth]{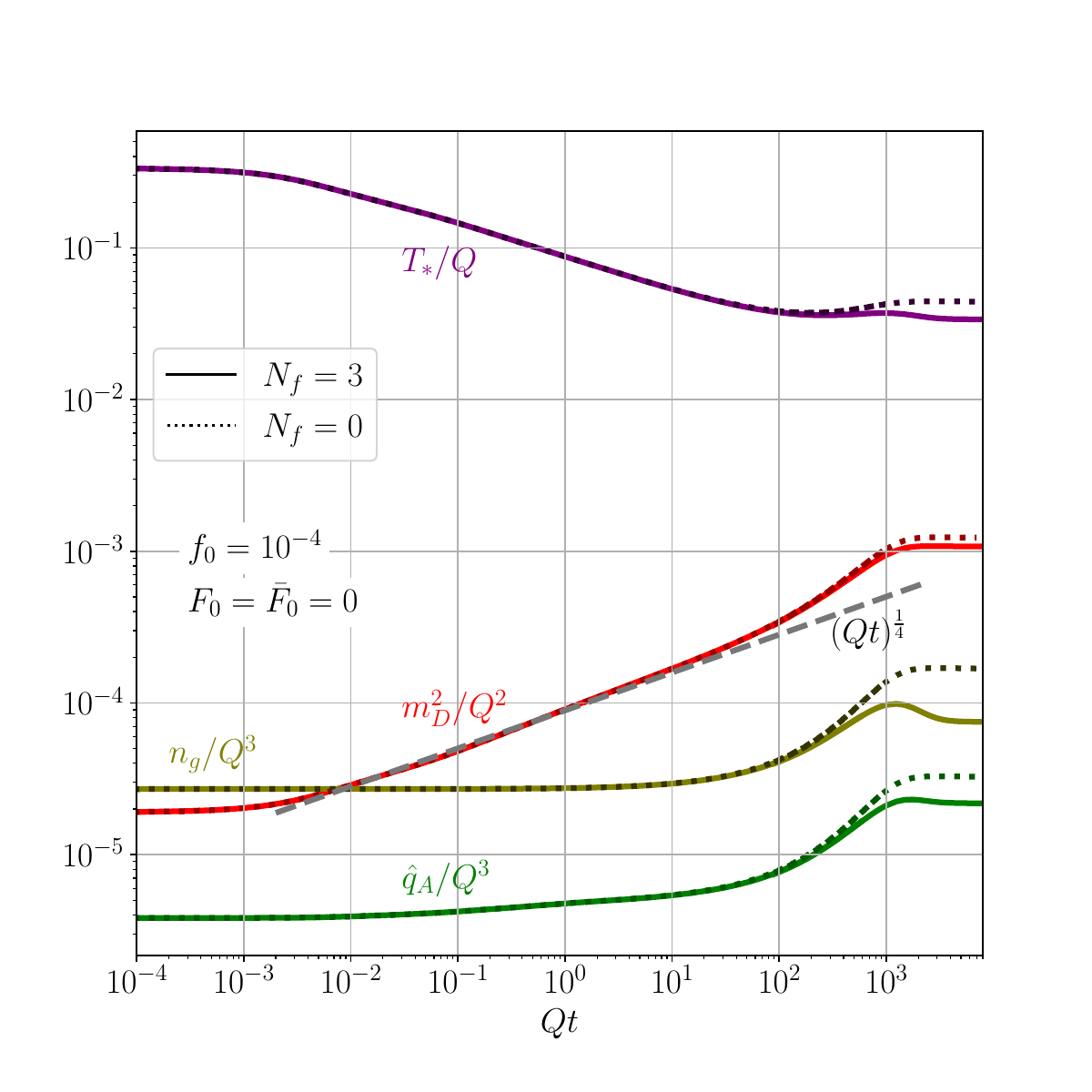}
\end{subfigure}
\begin{subfigure}[b]{0.32\textwidth}
	\centering
	\includegraphics[width=.95\textwidth]{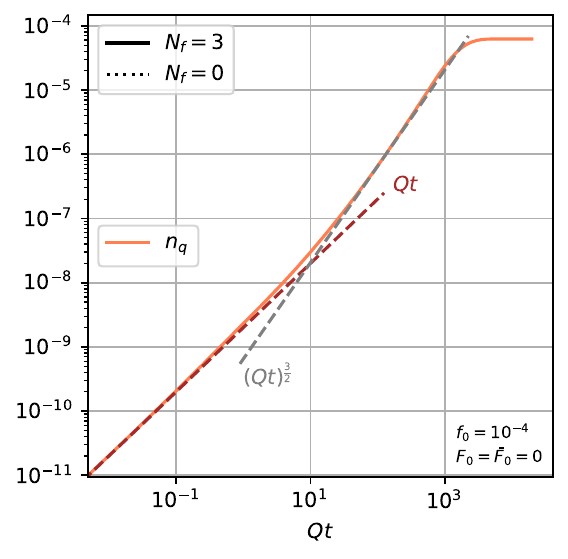}
\end{subfigure}

\caption{Parametric estimate for the underpopulated scenario (left) and numerical results for different macroscopic quantities (center) and quark number density (right) with $f_0=10^{-4}$.}
\label{Fig:underpp}
\end{figure}

First, the system will radiate a large amount of soft gluons which will fill a thermal distribution with effective temperature $T_*$, higher than the equilibrium one, $T$. All macroscopic quantities are dominated by the hard partons and their time dependence is parametrically negligible. In the next step, the screening mass starts to increase as it is dominated by the soft sector of gluons. Meanwhile, the quark production experiences a speed up as soft gluons take control of the $g \rightarrow q \bar{q}$ and $gg \rightarrow q \bar{q}$ processes while $T_*$ decreases and drops below $T$. Finally, in the last stage, $\hat{q}$ and $m_D^2$ are both dominated by soft partons, which form a thermalized quark-gluon plasma at temperature $T_*<T$. $T_*$ is then heated up again by quenching hard partons. The parametric evolution of these quantities as well as their numerical results are plotted in Figure \ref{Fig:underpp}.

\section{The overpopulated scenario}

\begin{figure}[t]
\centering
\begin{subfigure}[b]{0.32\textwidth}
	\centering
	\includegraphics[width=.95\textwidth]{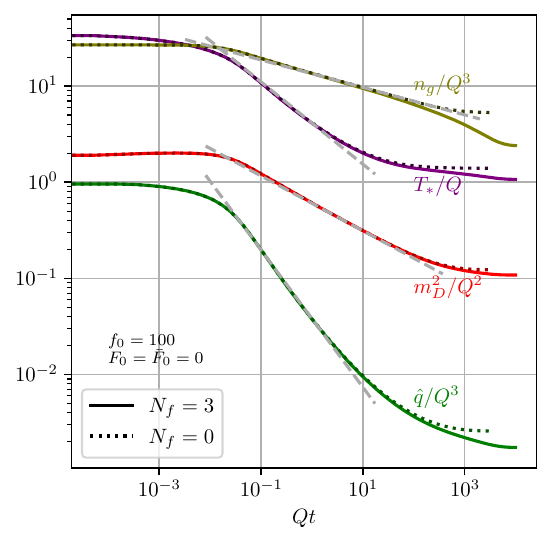}
\end{subfigure}
\begin{subfigure}[b]{0.32\textwidth}
	\centering
	\includegraphics[width=\textwidth]{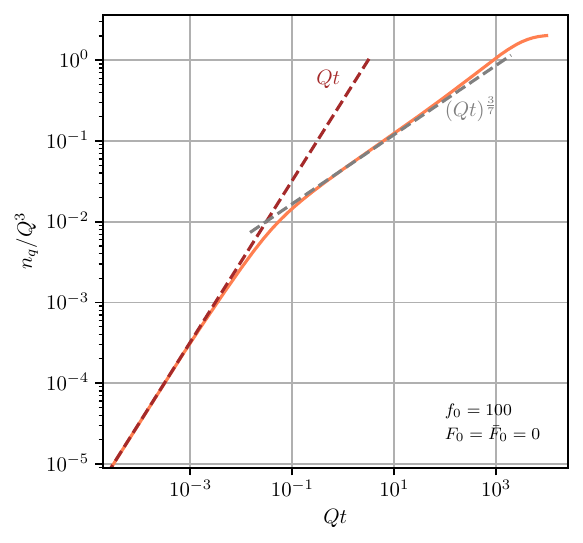}
\end{subfigure}
\begin{subfigure}[b]{0.32\textwidth}
	\centering
	\includegraphics[trim={10.2cm 0 0 10.1cm}, clip, width=0.92\textwidth]{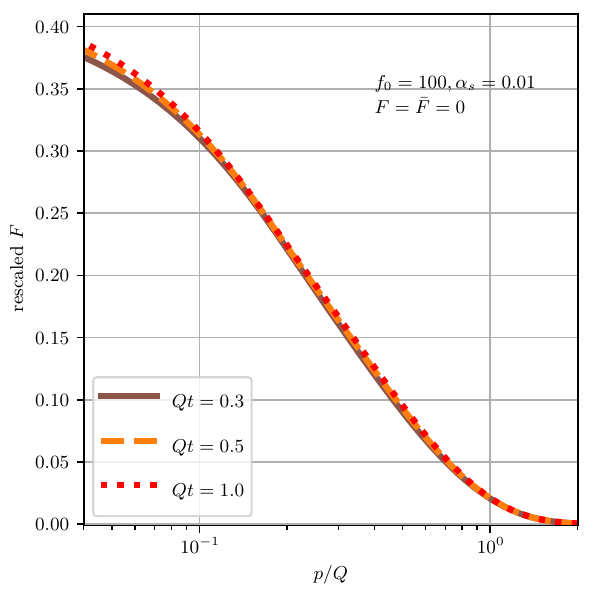}
\end{subfigure}

\caption{Numerical results for different macroscopic quantities (left) and quark number density (center) with $f_0=100$, and re-scaled quark distribution according with self-similar scaling(right).}
\label{Fig:overpp}
\end{figure}

Similarly to the previous case, the quarks do not play a major role in the parametric evolution of the system with respect with the pure gluon scenario. Thus, we only distinguish two different stages in the evolution \citep{BARRERACABODEVILA2022137491}. The first one is the rapid thermalization of the soft sector of gluons, and the second one the cooling of the system as shown in the left panel of Fig.~\ref{Fig:overpp}. Thermalization is complete when the effective temperature equals the equilibrium one.

The quark number density also increases linearly with time at the first stage. Nevertheless, in contrast with the underpopulated scenario, the quark production does not experience a speed up, but a slow down (see the center panel of Fig.~\ref{Fig:overpp}). This is due to the dominance of the momentum broadening in the second stage. The second stage shows the self-similar scaling behavior, as exemplified by the quark distributions in the right plot of Fig.~\ref{Fig:overpp} with $F = F_s((Qt)^{\frac{1}{7}}p / Q)$.

\section{Top-down thermalization}

At later times, and for not extremely underpopulated systems, we notice that the process $g \leftrightarrow gg$ establishes detailed balance before the system completely thermalizes. As a result, the gluon distribution achieves a thermal profile with a non-equilibrium temperature $T_*$. By the time this detailed balance appears, the gluon number is not constant, but decreasing as a result of the $g\rightarrow q \bar{q}$ and $gg \rightarrow q \bar{q}$ processes.

\begin{figure}
\centering
\begin{subfigure}[b]{0.32\textwidth}
	\centering
	\includegraphics[width=.95\textwidth]{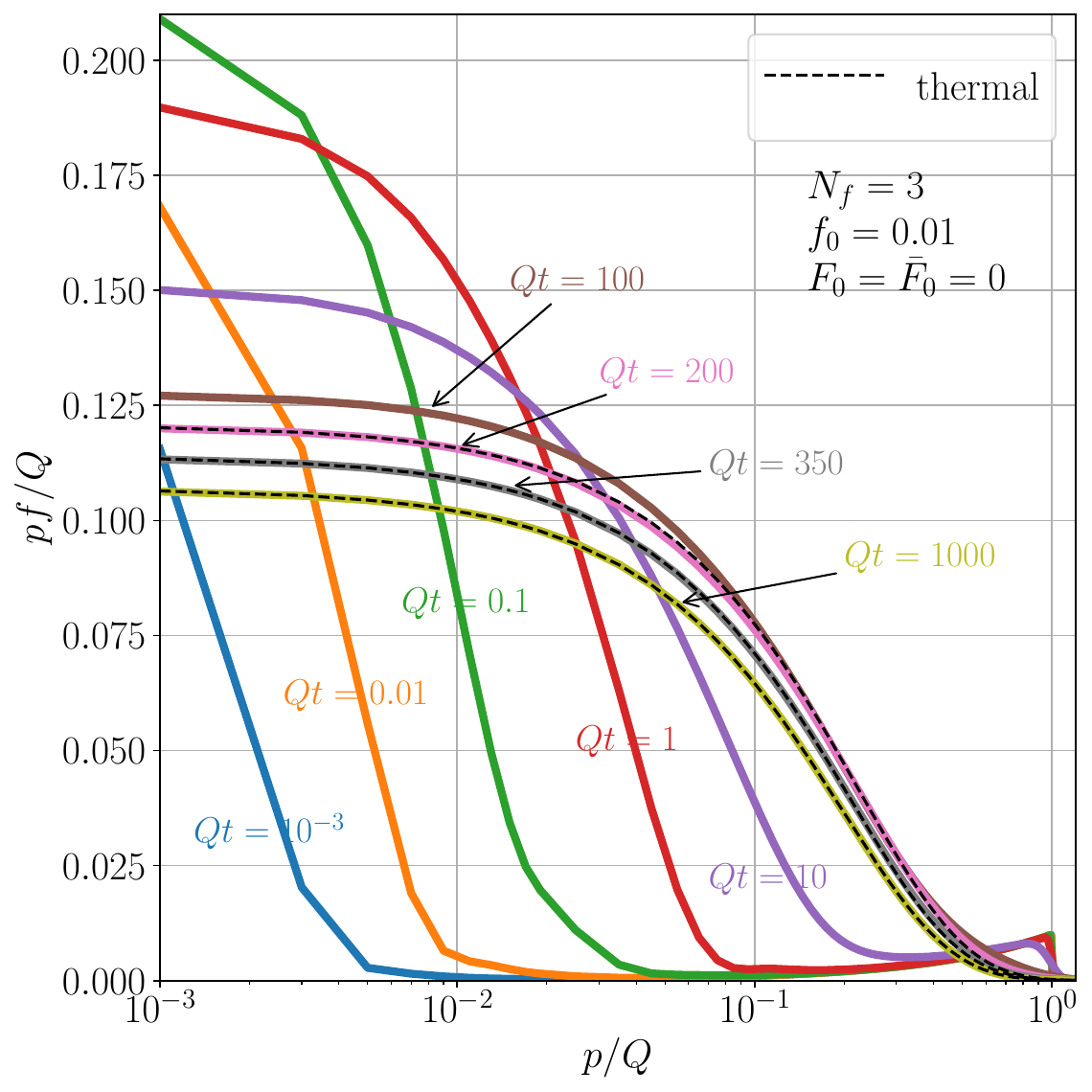}
\end{subfigure}
\begin{subfigure}[b]{0.32\textwidth}
	\centering
	\includegraphics[width=.985\textwidth]{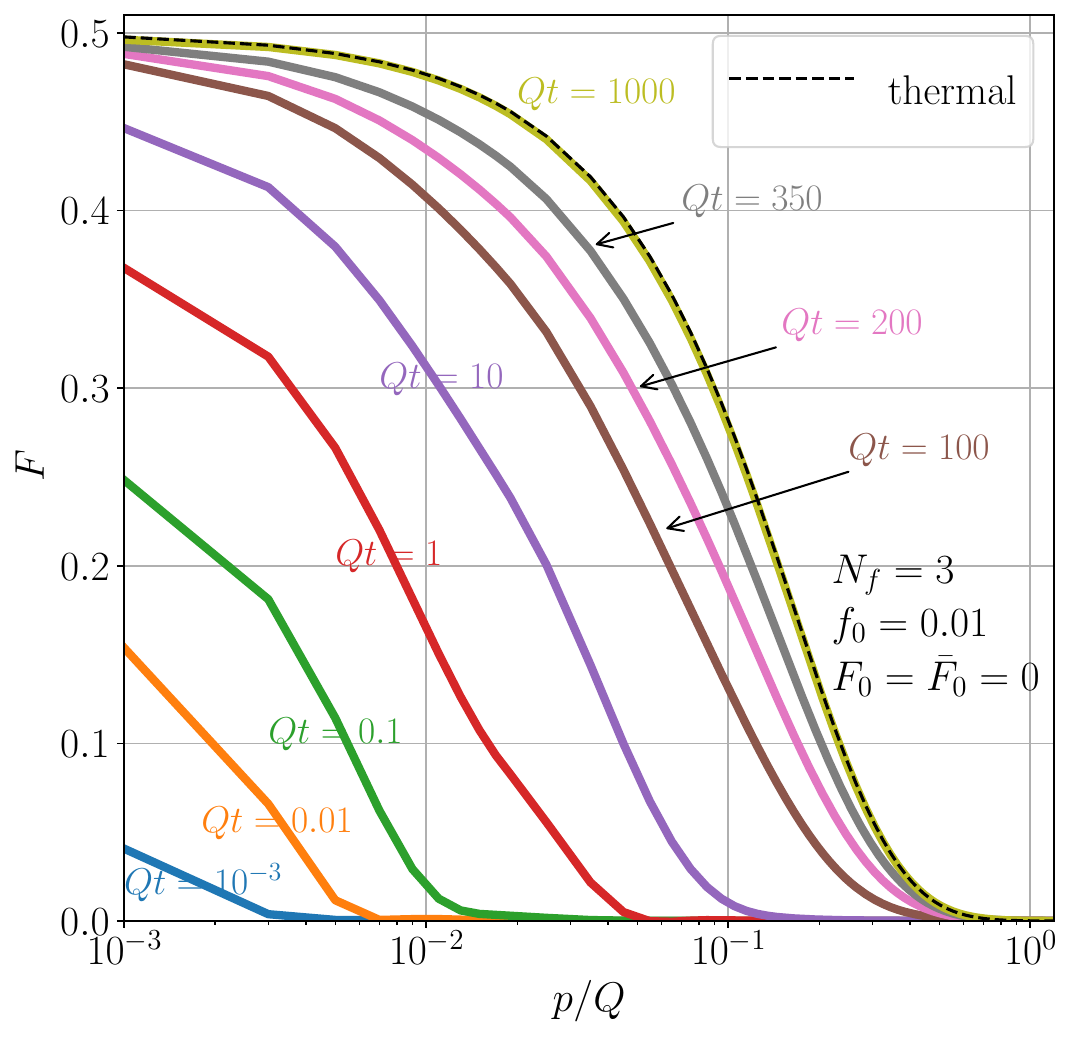}
\end{subfigure}

\caption{Gluon (left) and quark (right) distribution functions at different times.}
\label{Fig:topdown}
\end{figure}

As shown in Fig.~\ref{Fig:topdown}, the gluon subsystem achieves thermal equilibrium among itself before the quark distribution is able to fill a Fermi-Dirac profile, consistent with the results of \cite{Kurkela:2018xxd}. As a consequence, the thermalization time will be delayed in comparison to the pure gluon scenario. Also, such energy flow from gluons to quarks provokes a decrease of the effective temperature, such that the last stage of the thermalization follows a top-down fashion as long as the system is not extremely underpopulated.

%
%
%

\bibliography{bulk.bib}{}


\end{document}